# Effect of Information Technology on Job Creation to Support Economic: Case Studies of Graduates in Universities (2023-2024) of the KRG of Iraq

Azhi Kh. Bapir[1, 2], Ismail Y. Maolood[3, 4], Dana A Abdullah[3,4], Aso K. Ameen[3,4], Abdulhady Abas Abdullah[5]

[1] General Directorate of Sport-Ministry of Youth and Culture, Erbil 44001, F.R. Iraq
[2] Akre University for Applied Sciences, College of Administration and Economics, Department of Business Administration, F.R. Iraq.
[3] Department of Information and Communication Technology Center (ICTC) - System Information, Ministry of Higher Education and Scientific Research, Kurdistan Region – F.R. Iraq.
[4] Department of Computer Science, College of Science, Knowledge University, Erbil 44001, F.R. Iraq Iraq.
[5] Artificial Intelligence and Innovation Centre, University of Kurdistan Hewlêr Erbil, Kurdistan Region, Iraq

Select submission type:
☒Research Article (2500 – 6000 words)
_______________________________

Select submission Field (discipline):
☒ Engineering (Computer, Software, Electrical and Electronics, Civil, Mechanical, Manufacturing, Chemical, Petroleum, Geotechnical)
☒ Technology
_______________________________

Complete authors' information:
1st Author's organizational email: azhi.xalil@gmail.com
1st Author's ORCID link: https://orcid.org/0000-0003-1901-2776
1st Author's Google Scholar Citation link: https://scholar.google.com/citations?hl=enanduser=mqIZlEUAAAAJ

2st Author's organizational email: Ismail.maulood@mhe-krg.org
2st Author's ORCID link: https://orcid.org/0000-0003-1683-1493
2st Author's Google Scholar Citation link: https://scholar.google.com/citations?hl=enanduser=tdB8KHQAAAAJ

3st Author's organizational email: dana.ali@mhe-krg.org
3st Author's ORCID link: https://orcid.org/0009-0009-7610-3157
3st Author's Google Scholar Citation link: https://scholar.google.com/citations?hl=enanduser=w60L0LgAAAAJ

4st Author's organizational email: aso.khaleel@mhe-krg.org
4st Author's ORCID link: https://orcid.org/0000-0002-7037-7546
4st Author's Google Scholar Citation link: https://scholar.google.com/citations?hl=enanduser=3tL2kesAAAAJ

5st Author's organizational email: Abdulhady.abas@ukh.edu.krd
5st Author's ORCID link: https://orcid.org/0009-0007-5508-9371
5st Author's Google Scholar Citation link:
https://scholar.google.com/citations?user=fxIiXlMAAAAJ&hl=en&oi=ao

**ABSTRACT**

**The aim of this study is to assess the impact of information technology (IT) on university graduates in terms of employment development, which will aid in economic issues. This study uses a descriptive research methodology and a quantitative approach to understand variables. The focus of this study is to ascertain how graduates of Kurdistan regional universities might use IT to secure employment and significantly contribute to the nation's economic revival. The sample size was established by the use of judgmental sampling procedure and consisted of 314 people. The researcher prepared the questionnaire to collect data, and then SPSS statistical software, version 22, and Excel 2010 were used to modify, compile, and tabulate the results. The study's outcome showed that information technology is incredibly inventive, has a promising future, and makes life much easier for everyone. It also proved that a deep academic understanding of information technology and its constituent parts helps graduates of Kurdistan Regional University find suitable careers. More importantly, though, anyone looking for work or a means of support will find great benefit from possessing credentials and understanding of IT. The study's final finding was that information technology has actively advanced the country's economy. Not only is IT helping to boost youth employment, but it is also turning into a worthwhile investment for economic growth.**

*Index Terms—* Information Technology, Job Creation, Economic Growth, Graduates of Universities, Kurdistan Region of Iraq.

## I. INTRODUCTION

During the recent years the investments into the telecommunications sector have led to its further development and have become one of the essential sources of economic growth for many other important sectors and information technologies that form the society and information interchange(Goar et al., 2024). It can be stated that due to the impacts of the IT revolution, one of the fundamental paradigms of human development, or the process of expanding the choice in society to people, such as education, a healthy lifestyle, and living standards, is shifting (Yakunina and Bychkov, 2015). The issue is well known how innovations and technological development influences economic growth, as stated by (Lee et al., 2017). IT increases the informational capacity, establishes new lines of communication, redesigns work processing, and increases the efficiency of a spectrum of economic activities. Internet users are going on escalating throughout the world because of the fast expansion of the IT and telecom sectors, (Lee et al., 2017) especially that it has enabled young people to generate employment.

A single and perhaps one of the most outstanding trends of the last twenty years is the rapid and wide penetration of computers and information technologies in the workplace (Kabber et al., 2024). Consequently, there has been a great deal of hope that the emergence of IT can create wealth and growth and a lot of concern that the impact of IT on employment and skills has exacerbated the economic divide(Brynjolfsson and Hitt, 2000, Ameen et al., 2024b). While there is continuing debate and therefore still some confusion about what it all means and whether IT is indeed the central element of an information revolution that will fundamentally impact employment, earnings, well-being, the economy, and society generally (Brynjolfsson and Hitt, 2000)There is no doubt that the role of IT has been central to change processes in organizations and businesses.

Additionally, Information technology has an early impact on small businesses, which will strengthen the economy and create jobs (Bapir, 2023). Like other countries, the technology quickly reached the Kurdistan Region of Iraq (KRI) and divided the Kurdistan community into two parts, some of which were able to benefit from technology, and some not only could not benefit from it, but it had become a major disaster for them(Yahia and Miran, 2022, Abdullah et al., 2024). KRI condition is good for economic development for this reason; it has the potential to be a regional economic powerhouse because of the proven regular resources and large population of working age force and most importantly the fact that IT has contributed in sector enhancement (Bapir, 2023). Moreover, there is a large number of graduated people in KR each year who are unable to get employed (Bapir, 2023, Orsam, 2021), however, in most developed countries IT emerges as the major prospects for employment.

The purpose of this study was to find out how much graduates from universities can take advantage of the skills of information technology to get job opportunities. Therefore, this research is planned to discover how job creation can be provided through the effectiveness of IT, which further brightens the country's economy. When many graduates have been unable to take advantage of digital skills to get rid of unemployment. The main research question of the study was to ask about the impact of IT on finding job opportunities for people who own degrees, which would take economic development to a new stage.

## II. LITERATURE REVIEW

*I. A Thorough Analysis of Information Technology*

Over the past few decades, the field of information technology (IT) has grown rapidly, particularly with the introduction of the internet in the 1990s. Also, the world is now interconnected within the cloud due to the internet's rapid advancement (Rocha et al., 2024). Advances in IT have led to the use of software as a service, cloud computing, secure databases, platforms as a service, and many more. The efficiency and management of security and storage systems have increased, and computers can now operate and maintain IT services(Rocha et al., 2024).

The definition of information technology given is "any activity that involves information processing as well as integrated communication using electronic equipment" (Victoria, 2020). This phrase also encompasses all ITs irrespective of their application in information systems, industrial process control, communication between two computerized companies, or individuals' utilization of computational devices. According to (Burgelman, 2001), IT could be defined primarily in terms of the assets employed in the transformation and handling of a firm's information. Such equipment includes hardware, software, communications (voice, data, and video), and the personnel that come with them.

Information technology is defined as the roles that the computers, the software and the communications systems might assume(Burgelman, 2001). According to (Victoria, 2020), there are many types of IT, but understanding the set that IT is made of is essential to get references on potential strategic uses of IT. It is possible to classify the following as IT categories:

1) Information systems
2) Hardware technology
3) Office automation
4) Computer engineering and design
5) Industrial automation
6) Specific automation resources
7) Multimedia resources

The subject is crucial to classify information technology since any type of classification quickly becomes dated due to the high rates of development in this sphere; therefore, this systematization of the most relevant set of information technology provides a quick reference for investing in the primary strategic application areas. As explained by (Ameen et al., 2024a), the following IT typology and illustrations:

1) Systems development technologies, such as project management procedures, methodologies for testing and debugging software, methodologies for designing databases, and methodologies for programming.
2) Information technology planning-related technologies, such as data process modeling, the informatics master plan, and information technology methodology.
3) Technology for manufacturing operations and processes, such as PCP, capacity planning, and performance management.
4) Application software support technologies include operating systems, database management systems, teleprocessing software, utilities, performance monitors, programming languages and application generators. The further step to be undertaken after being familiar with the various types of currently used IT is to know how these types of IT can be supportive of organizational strategies in an organization(de la Garza Montemayor et al., 2021). Information technology is going to be around for a long time, and one of the most fascinating things about it is how fast it is developing. With the advent of the internet and smartphones, technology has completely changed the way we work and live, and it does not seem to be stopping anytime soon (Arai, 2023). Here are some clarifications:

   a) Artificial Intelligence (AI) and Machine Learning (ML).
   AI and Machine learning are two of the largest advancements that one can look at under Information Technology. AI is defined as the development of computer systems capable of performing tasks that otherwise require a human touch. Decision-making, pattern recognition, and understanding natural languages are just but some of the capabilities.

   b) Cloud Computing
   Another development that is altering how we utilize technology is cloud computing. It defines the Internet-based delivery of computer utilities including hosting, data warehouses, computing, connections, application, and intelligence. Several businesses can get access to the powerful tools and applications through cloud computing model; however, the companies do not need to purchase and maintain their own equipment. They are no longer required to maintain their own infrastructure as a result. For businesses, this means that managing, storing, and analyzing enormous volumes of data is now simpler and less expensive.

c) 5G and Edge Computing

The internet will operate differently for us thanks to two trends: 5G and edge computing. The fifth generation of mobile networks, or 5G, is expected to provide quicker and more dependable internet access. When data is processed near, its source as opposed to centrally, it is referred to as edge computing.

d) Virtual Reality (VR) and Augmented Reality (AR)

Our interactions with the world are changing as a result of emerging technologies like augmented reality and virtual reality. In virtual reality (VR), a fully immersive simulated environment is created for people to explore. A technique called augmented reality (AR) blends digital data with the real environment. This makes it possible for users to engage with virtual items more naturally. Applications for these technologies are many and include everything from training and education to gaming and entertainment. Virtual reality (VR) and augmented reality (AR) have the potential to revolutionize how we perceive the world in the future.

e) Digital Information

Ultimately, the most significant factor influencing how information technology will develop in the future is digital transformation. New products and technologies are being developed as a result of the digital transformation. The way we live and work is being affected by these changes. Businesses are being presented with opportunities by developments in cloud computing and artificial intelligence. They can supply new goods and services to clients, automate operations, and collect and analyze data in real-time. To sum up,(Adhikari, 2023) says that information technology is influencing the course of the future. It is changing how we communicate, work, and live. Over the coming years, 5G, edge computing, AL, and ML will all have a greater impact on our daily lives. Our daily routines will be more impacted by these improvements. Working in IT is thrilling these days. In the upcoming years, there will likely be advancement that is a lot more fascinating.

B. *Information Technology's Influence on the Growth of Job Creation*

Technology is a two-edged sword that both empowers and restrains people. It also has the potential to be both beneficial and harmful for the economy, jobs, and employment(Peng et al.,

2018). During the time of the invention of computers, some researchers initially showed that computers will destroy job opportunities, but after research and the introduction of computers into the work process, most supported that computers can advance people in many areas and provide many job opportunities(Attewell and Rule, 1984).

Concerning (Clarke, 2022), the idea of work and occupations have been empowered through technological advancement. Due to recent advancement in technology, there are certain jobs that people use to do but can now be done by machines. As a result, there is less demand for human labor and an increase in productivity and efficiency at work, especially for people with advanced degrees, professionals, and other highly qualified individuals who can benefit most from the information technology age (Clarke, 2022). Some people get worried that as technology is being applied then there would be little demand for people, although not normally true. This feat can be said to be logical because as technology advances some occupations may be rendered irrelevant, but new occupations are also defined to fill the new technological setups(Arsic, 2020). Worldwide, the opportunities that could be robotized based on one early prognosis(Manyika, 2017).

However, discounting the historical lesson that new technology results in the disappearance of certain jobs but creates totally new jobs is counter-productive; the "jobs or no jobs" criterion can be misleading and is actually a case of "false positives"(Restrepo, 2018). Think in terms of particular jobs within given professions shifting to the mechanical (i.e., technology as a collaborator not a substitute) (or rather technology as replacing certain functions with individuals). One should take into account the totality of possible automations of the given activity: the range and the degree. The belief is that advanced technology will substitute human labor in all sectors of the economy, even when people have more job training(Restrepo, 2018). Therefore, instead of occupations, some duties may be replaced. However, there are other effects well captured by offsets originating from technology that actually raise employment, such as embodied technology in higher capital accumulation or creation of new jobs that require more labor and less machinery(Autor, 2015).

Furthermore, (Bresnahan et al., 2002) argue that there is another problem of misconception of the effects of technology on labor: Misimetrized unemployment belief, that is, all workers within a given industry or occupation will suffer from unemployment in equal measures and at the same time. This projection uses a low power lens to observe an intricate, temporally arising process. Certainly, there will always be tasks and jobs that will prove to be very immune to change instigated by technology. This area contains atypical structural and functional activities. When the cost of producing a subset of jobs is reduced, concerns with the comparative advantage of labor for various tasks emerge; automation escalates the demand for labor in other related tasks (Bresnahan et al., 2002).

Another area of technology pointed out by (Leso et al., 2018) to have tension and expected to digitize occupations in the future is Internet of Things or the integration of objects and devices. As this technology can enhance productivity and effectiveness in the workplace, at home and recreationally it can revolutionize the ways in which all aspects of living are performed. Take for example a factory where the machines themselves are wired and can communicate with each other to accomplish some given job. This would relatively enhance the efficiency of the manufacturing process and consequently the demand for human labor (Leso et al., 2018).

(Clarke, 2022) elucidates that the internet of things has the ability to develop brand-new employment categories that do not even exist now. Businesses will require personnel, for instance, to handle and analyze the data that is gathered by all of these linked devices. Can anticipate a wide range of career opportunities being produced as a result of the internet of things as it spreads. The demand for IT consulting services has increased as much as the number of available IT jobs. There is a demand for people to work because many people need assistance using all these new technologies(Clarke, 2022).

Also,(Zhao et al., 2022) state that a promising new sector that has the potential to revolutionize the world economy is IT. New avenues for economic expansion in underdeveloped nations are being made possible by digital technologies; in addition, the development of the internet is a keystone and catalyst for increased global connectivity and economic revolution(Zhao et al., 2022). The best example is Africa, which has been able to create the most jobs for university graduates through information technology while increasing annual GDP growth by about 1.5 percentage points and reducing the number of poverty heads by 0.7 percentage points per year(Wankuru, 2019).

Georgescu in 2021 refer that the majority of the highly skilled labor force could potentially be employed in the IT sector. So, it is clear that the country needs to invest in education and/or skill development if it hopes to attract young people to the lucrative job growth in the IT sector (Georgescu Irina, 2021). If their education and skill levels stay at secondary or lower levels, and basic skills, social skills, and problem-solving abilities are among the crucial competencies required in the IT business (Georgescu Irina, 2021). Therefore, the development of jobs in the industry should go hand in hand with the ability to fill identified skills gaps as well as current talents.

*C. Information Technology Has an Impact on the Economy's Growth*

After the first electronic computer was created during World War II, the digital and technical revolution began. Over the following ten years, both data storage and processing speeds increased significantly. In the 1980s, homes, enterprises, research facilities, military services, and public administration all adopted computer use(Arsic, 2020). Concerning (Faucheux et al., 2010), technology is viewed as a growth driver for the economy. It has an impact on it in two ways: directly through the industries that produce ICT and directly through the industries that are users. All economic sectors now employ ICT to the point where, in developed economies, indirect productivity gains associated with digitization and how it is used are frequently considered as the key vector of growth (Faucheux et al., 2010).

The COVID-19 epidemic has boosted technological and digital development since it has increased motivation for online labor (Bashir et al., 2022). COVID-19 has accelerated digital progress in corporations as well as in governmental and personal settings. Nonetheless, during the COVID-19 epidemic, countries were able to continue operating normally thanks to digital and technological advancements (Amankwah-Amoah et al., 2021). In demonstrated that encouraging the growth of IT has significant potential to inspire digital and technological advancement in all spheres of society, the economy, and rising GDP. The standard of living, productivity, and employment rate have all increased because of technological advancement. It improves productivity, economic expansion, and people's access to jobs, financing, and education (Zhao et al., 2022).

For the years 1995 to 2000, France had annual contributions from IT of 0.35% and equipment and buildings of 0.52%. (Mefteh and Benhassen, 2015). Collecchia and Schreyer findings for France are very comparable to their findings for Germany, Italy, and Japan. Consequently, across the four periods from 1980 to 2000, IT contributed between 0.20 and 0.40 percent annually to economic growth in these three countries (Colecchia and Schreyer, 2001). The biggest impact was seen in the United States with rates being double those recorded in France. It varied from 0.45% per year up to 1995 and even 0.90% per year in the period of 1995-2000. Australia, Canada, Finland and United Kingdom have all been way behind United States. These nations are in between the first group of nations (Germany, France, Japan, and Italy) and the United States (Colecchia and Schreyer, 2001).

Additionally, an examination of the work by Oliner and Sickel for the United States and Cette, Mairesse, and Kocoglu for France reveals that the IT sector's contribution, which was 0.3% per year in the United States until the middle of 1995, has lately increased to 0.8%. In the same time frame, it increased in France from 0% to 6% annually. Consequently, these findings lead to a very important conclusion about the IT-using industries throughout economic. IT use appears to be more significant in the United States. The above authors concur that IT grew in the late 1990s and subsequently enhanced the contribution of economic growth (Oliner, 2000, Oliner and Sichel, 2003, Cette et al., 2002).

### III. METHODOLOGY

The research employed descriptive research approach within a quantitative research paradigm to analyze variables. This research concentrates on graduates of universities in the Iraqi Kurdistan Region provinces of Erbil, Sulaymaniyah, and Duhok (see Table I). This is followed by purposive sampling that was used to select a sample size of 314 people, because the potential audience size was huge. In easy terms, non- probability technique involves purposive sampling, in which the researcher picks people from the population that the researcher thinks would be capable of providing the required or correct information (Etikan and Bala, 2017). The method is also often commended for its validity, reliability and non-bias nature that the current study has adopted. In an effort to gather the information, the investigator developed a questionnaire. The main advantages of using a questionnaire to gather information would be the low overall cost of the project and timesaving, respondents are able to provide honest and, ultimately, anonymous answers to the questions posed, which makes it easier to obtain the basic facts and data. The researcher employed a pretest in a bid to isolate the effect of possible questionnaire errors and to determine the possible the time necessary to complete the instrument checking the validity of questions is an important aspect of the pre-testing, whereby the questionnaire is used. Some of the respondent's pre-test the questionnaire to eliminate any irrelevant items in a sample of respondents. Roberts-Lombard noted that even the language and sequence of a survey and the overall design of the questions must be put in analysis. In an effort to design the survey, a closed-end questionnaire was adopted (Roberts-Lombard, 2002).

TABLE I
NUMBER OF PARTICIPANTS GRADUATES OF UNIVERSITIES IN KURDISTAN REGION GOVERNMENT-IRAQ

| Number of participants graduates | Name of Universities | Location |
| --- | --- | --- |
| 26 | University of Teshk | Erbil Province |
| 48 | University of Knowledge | Erbil Province |
| 62 | University of Salahaddin | Erbil Province |
| 27 | University of Duhok | Duhok Province |
| 13 | University of Kurdistan | Erbil Province |
| 15 | University of Soran | Soran Administrative Boundary |
| 47 | Erbil Polytechnic University | Erbil Province |
| 31 | University of Jehan | Erbil Province |
| 18 | University of Sulaymaniyah | Sulaymaniyah Province |
| 12 | University of Akre | Duhok Province |
| 5 | University of Rapareen | Rapareen Administrative Boundary |
| 6 | University of Garmeian | Garmeian Administrative Boundary |
| 4 | University of Zaxo | Zaxo Administrative Boundary |

## IV.  RESULT AND DISCUSSION

*A.  Demographic Outcomes*

This section concentrates on the study participants' gender, age, type of working, and educational degree levels. This is important because it shows that the people who provided the data were the ones with the closest relationships. The findings show that our investigation yielded nearly all of the perfect questionnaires for this kind of research.

According to Fig. 1(a), 40% of study participants were women and 60% of participants were men. This makes sense given that men dominate the workforce, the educational system, and other areas of society in the Kurdistan region of Iraq (KRI), and Fig. 1(b).  Shows that most of our respondents, however, were in the age range of 18 to 25. This is due to the fact that the study focused mostly on young people, who are both essential to the growth of the country's economy and the group most impacted by unemployment. Yong adults between the ages of 18 and 25 typically inquire about jobs or job openings in KRI.

Regarding (Bapir, 2023), a large number of Kurdistan Region (KR) institute and college graduates are produced each year. Upon graduation, they are eager to work in both the public and private sectors; if they are unemployed, they will attempt to start their own firms. Additionally, the KR is home to several sizable businesses that are eager to hire, particularly for individuals with college degrees and fluency in other languages(Bapir, 2023, Orsam, 2021).

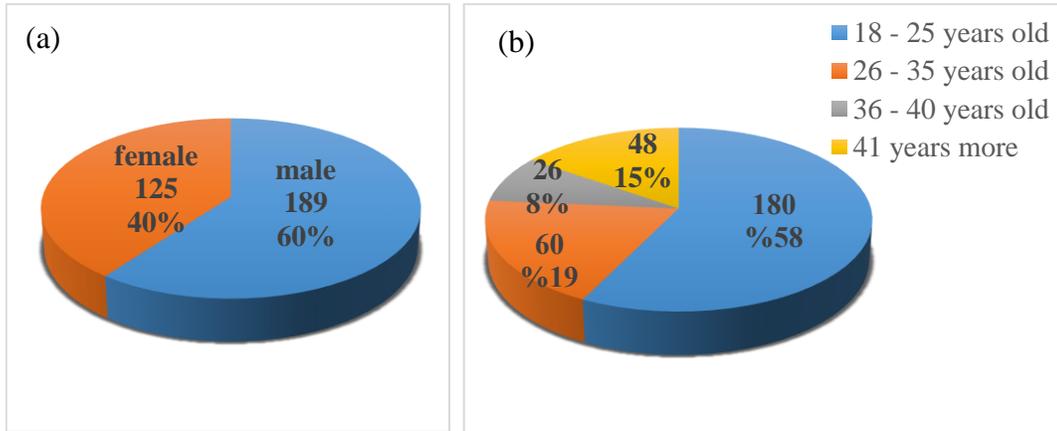

Fig.1. (a) Pie-chart illustration of gender distribution, and (b) shows the respondents' age distribution

Fig. 2(a). Therefore demonstrates that the majority of research participants hold bachelor's degrees and institute certifications. In comparison to other levels, the bachelor's and institute degree included the highest percentages; of the 314 participants, 160 have institute degrees, and 68 are college graduates.

As reported by Al-Mihya in 2017, the Kurdistan Region experienced a severe financial and economic crisis after 2014, which resulted in a large number of job losses but also the bankruptcy of numerous banks and individual businesses. Furthermore, the Kurdistan Regional Government's (KRG) market continued to contract year after year, rendering employment for the majority of people unattainable (Al-Mihya, 2017). For this reason, Fig. 2(b). demonstrates that, of the 314 participants, 152 are graduates of the department of information technology who are unemployed or have not been able to obtain employment; only 65 are employed by the government, 40 by the private sector, 57 have jobs outside of the public and private sector.

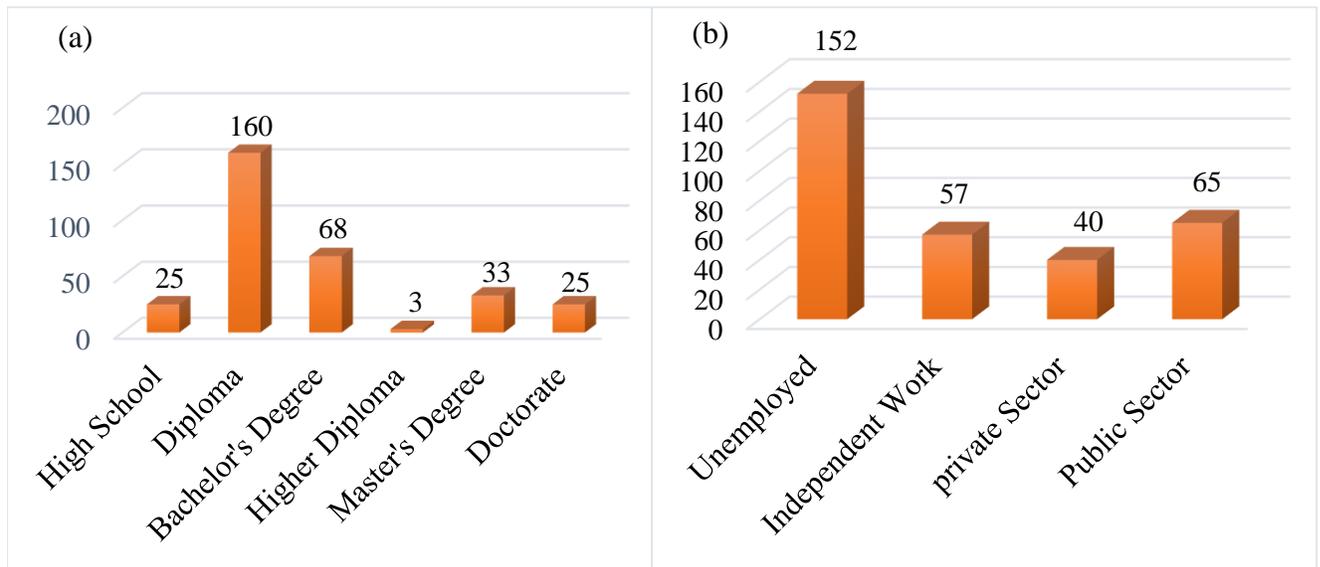

Fig.2. (a) Educational Qualifications of Research Participants, and (b) Employment Status of IT Graduates

*B. Research Findings*

The following section displays the results of the research questions that are the respondents' views on the definition and future of information technology, as well as finding jobs and developing the economic sector by IT. These results will be useful for graduates of universities in the Iraqi Kurdistan Region if they are able to keep learning new information technology skills. These findings specifically demonstrate that university graduates from the Iraqi Kurdistan Region

can get exceptional job opportunities through IT. They also reveal how information technology

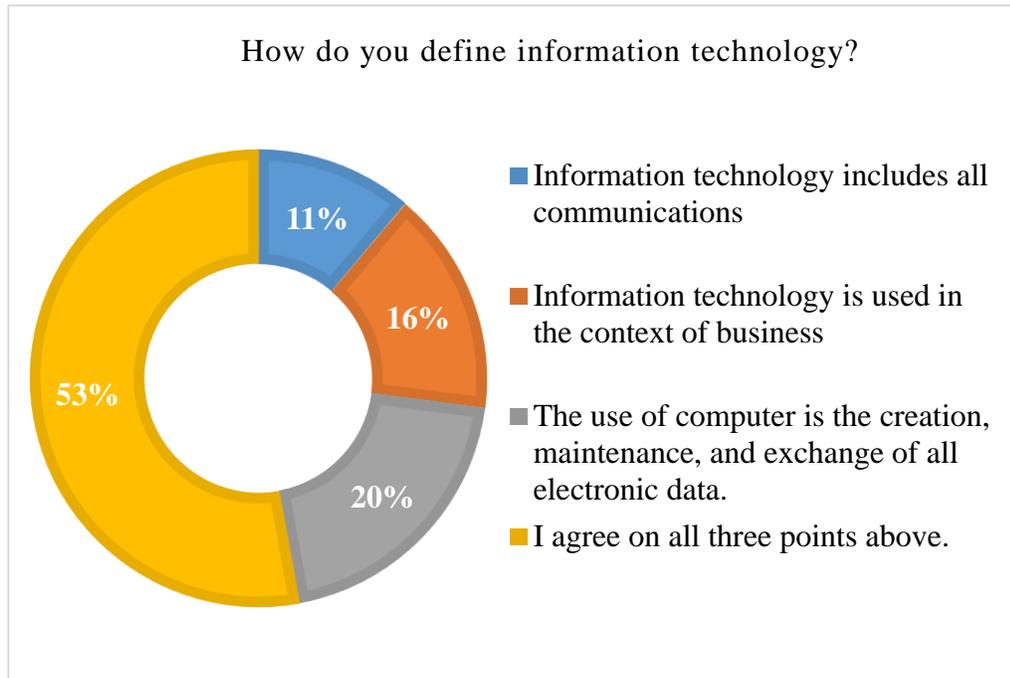

plays a major role in enhancing the national economy.

Fig. 3. A Thorough Analysis of Information Technology

According to (Victoria, 2020), information Technology is defined as "any activity involving information processing and integrated communication through electronic c equipment". This phrase is more general and includes all IT developments for information systems, industrial processes, inter-company or intra-organization communication of two companies or organization via computers and even using computational resources for personal purposes. In Fig.3, the majority of respondents concur on the fourth point, which states that information technology encompasses all forms of communication, is utilized in corporate settings, and that using a computer involves the generation, upkeep, and exchange of all electronic data.

(Adhikari, 2023) refers to IT influencing the course of the future. It is changing how we communicate, work, and live. Over the coming years, 5G, edge computing, AL, and ML will all have a greater impact on our daily lives. Our daily routines will be more impacted by these improvements. Working in IT is thrilling these days. In the upcoming years, there will likely be a lot more advancement that is fascinating. As shown in the Fig.4 (a), out of 314 respondents, the majority (172 of them) think that information technology will have a bright future and a broad use of it. Twenty-seven people do not have an opinion about information technology, 79 believe that the future is not for all fields, and 36 believe that information technology will not have a bright future.

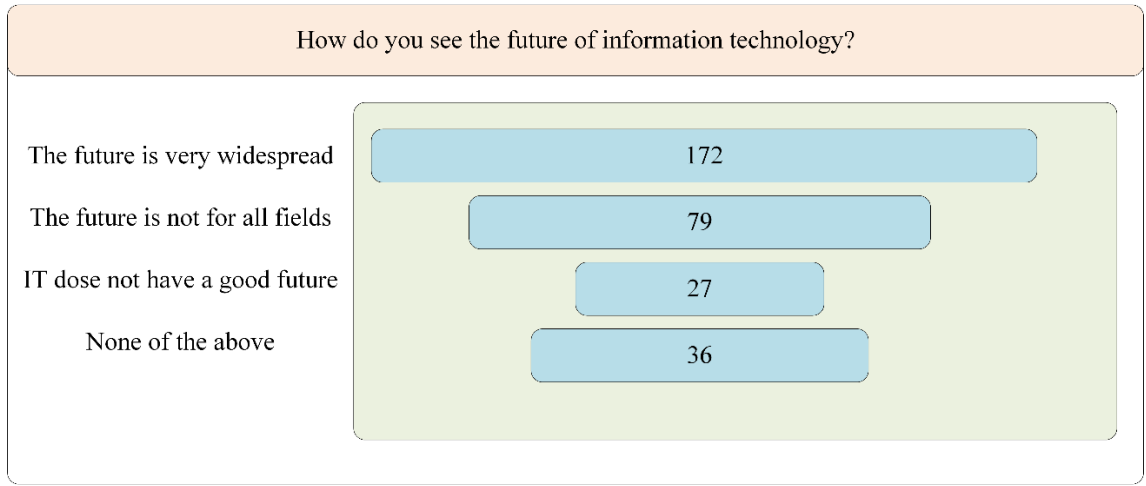

Fig. 4 (a). Thorough Analysis of future of information technology

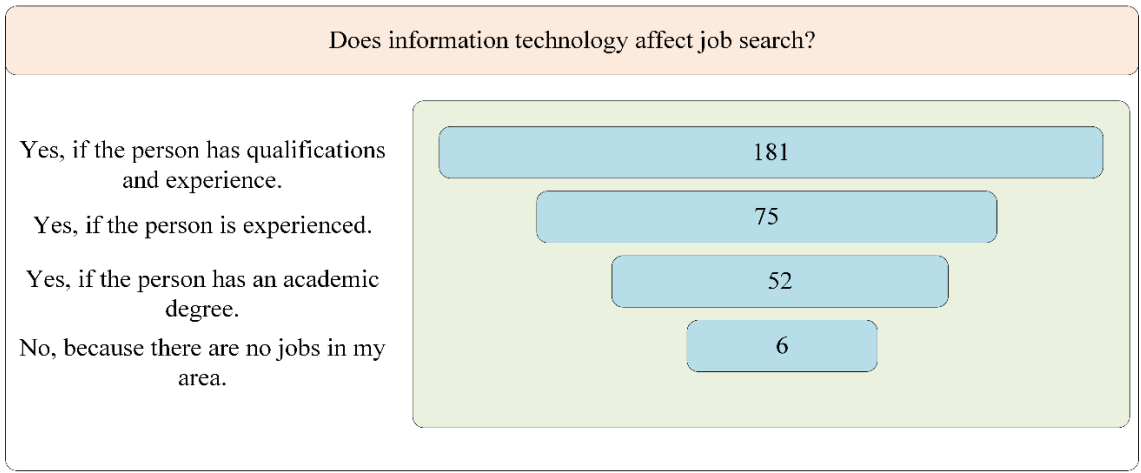

Fig. 4(b). Information technology influence on the growth of job creation

As mentioned by Clark, in the year 2022, it has referred to the future of work and occupations and the advantages of technology. There are many professions that were previously accomplished by people and are now practically handled by computers due to present developments. Especially people with advanced degrees, professionals, and other highly qualified individuals can benefit most from the IT age (Clarke, 2022).

Besides, (Arsic, 2020) noted that people used to state that with advancement of technology,

machines and turn into non-existent can replace certain professions, however, fresh professions will be generated. For this reason, as shown in Fig. 4(b), the majority of respondents think that using information technology to get a job is a far better option for someone with experience and qualifications. So out of 314 participants, 181 say (yes, if the person has qualifications and experience), 52 say (yes, if the person is certified), 75 say (yes, if the person is experienced), and 6 say (yes, if the person is experienced).

According to(Peng et al., 2018), technology is a two-edged sword that both empowers and restrains people. It also has the potential to be both beneficial and harmful for the economy, jobs, and employment.

Additionally, (Attewell and Rule, 1984) explained that during the time of the invention of computers, some researchers initially showed that computers will destroy job opportunities, but after research and the introduction of computers into the work process, most supported that computers can advance people in many areas and provide many job opportunities. In the light of Fig. 5. First Phrase, most respondents strongly agree that technology is a double-edged sword that both empowers and controls people; especially IT has a great impact on job creation. That is, 95 participants responded with an agree, 170 participants responded with strongly agree, 23 participants responded with disagree, 10 participants responded with strongly disagree, and 16 participants responded with not decided.

Concerning (Faucheux et al., 2010), technology is viewed as a growth driver for the economy. It has an impact on it in two ways: directly through the industries that produce information and communication technology (Victoria) and directly through the industries that are users. All economic sectors now employ ICT to the point where, in developed economies, indirect productivity gains associated with digitization and how it is used are frequently considered as the key vector of growth(Faucheux et al., 2010). Based on Fig. 5. Second Phrase, the majority of respondents strongly agree that IT will be a major factor in economic recovery, especially in countries where information technology is very successful and has even been able to influence the development of the economic sector. Accordingly, out of 314, 84 respondents said they agreed, 181 said they strongly agreed, 18 said they disagreed, 14 said they strongly disagreed, and 17 said they were unsure.

As mentioned through(Mshvidobadze and Ginta, 2022), encouraging the growth of IT has significant potential to inspire digital and technological advancement in all spheres of society, the economy, and rising GDP. The standard of living, productivity, and employment rate have all increased because of technological advancement. It improves productivity, economic expansion, and people's access to jobs, financing, and education(Mshvidobadze and Ginta, 2022).

Besides, (Mefteh and Benhassen, 2015) clarify: For the years 1995 to 2000, France had annual contributions from IT of 0.35% and equipment and buildings of 0.52%(Mefteh and Benhassen, 2015). Collecchia and Schreyer's (2001) findings for France are very comparable to their findings for Germany, Italy, and Japan. Consequently, across the four periods from 1980 to 2000, IT contributed between 0.20 and 0.40 percent annually to economic growth in these three countries (Colecchia and Schreyer, 2001)

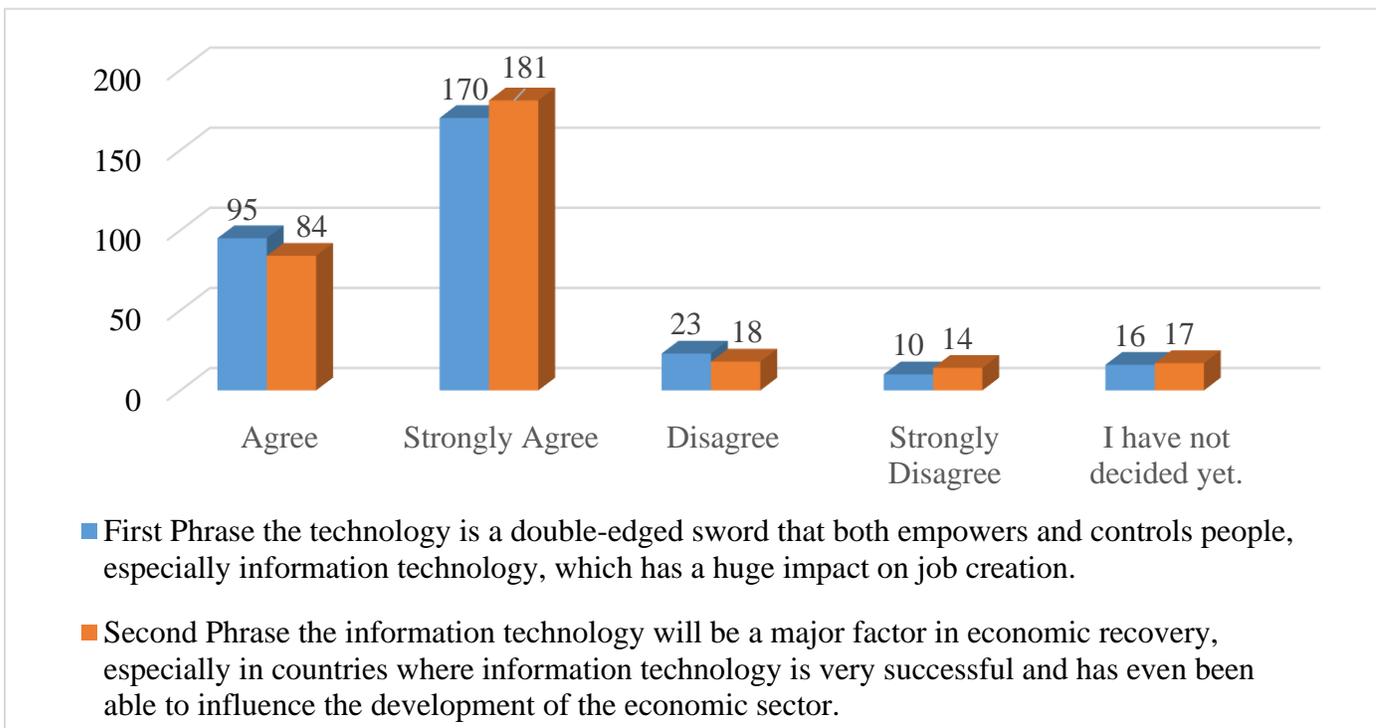

Fig. 5. Information Technology Has an Impact on the Economy's Growth

The highest increase was observed in the USA; moreover, it was two times higher than in France. It varied between 0.45% per annum up to 1995 and even 0.90% per annum for the period from 1995 to 2000. Australia, Canada, Finland and the United Kingdom have all done worse than the United States. These nations are somewhere on the middle concerning the first group of nations (Germany, France, Japan, and Italy) and the United States(Colecchia and Schreyer, 2001). Thus, according to the **Fig. 6**, the majority of the 314 participants believe that information technology has the potential to boost the economic sector in your country. Accordingly, 110 say: Yes, if the government has a regulatory program on how to use it; 116 say: Yes, because many college graduates of IT departments have entered the labor market in the past three years. 66 say: No, because the quality of the universities in the Kurdistan Region is low and they do not have the capacity to educate students in this field; 14 say: No, because the level of intelligence of the people in my area is weak; and 8 say: They are all true.

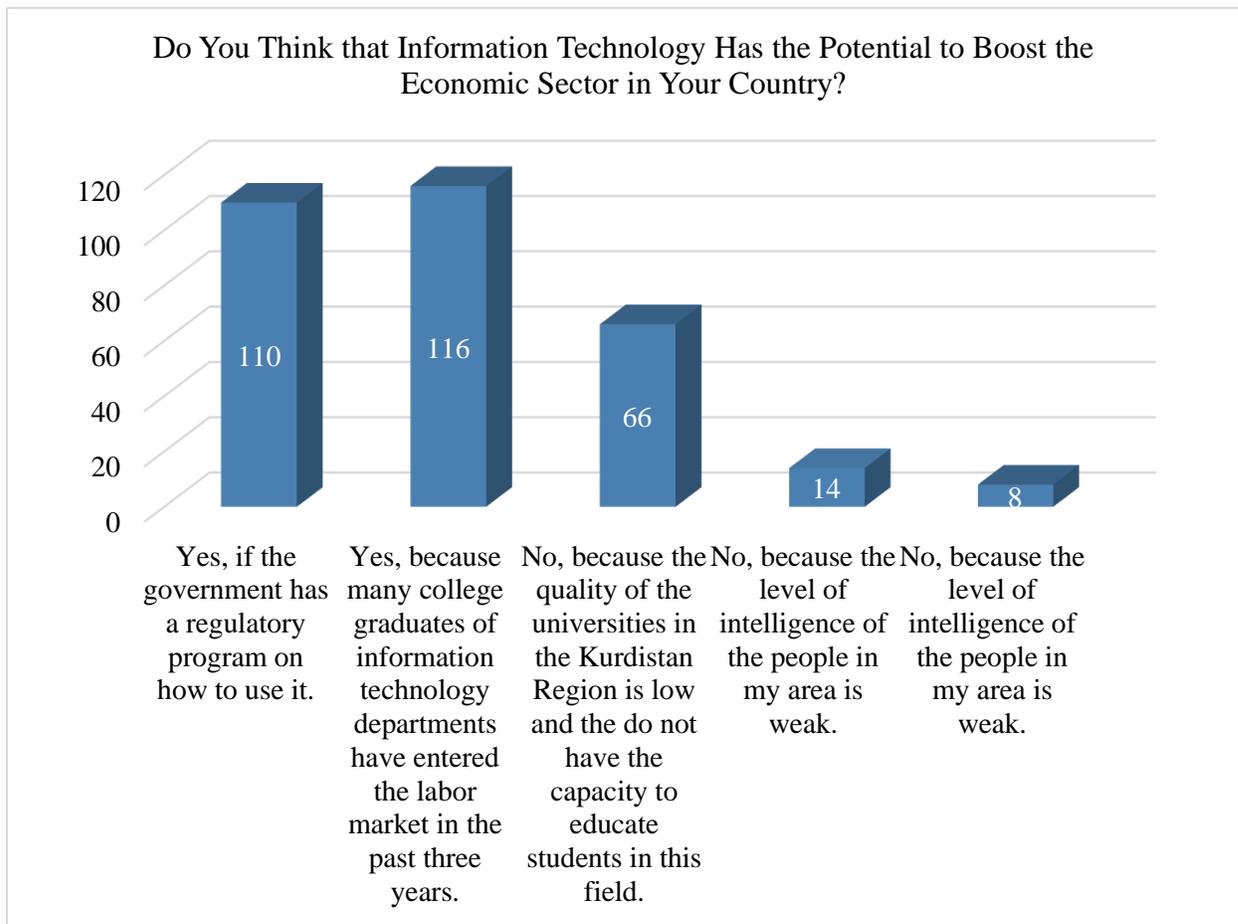

## V. CONCLUSION, IMPLICATION AND LIMITATIONS

The aim of this study is to evaluate how information technology can affect university graduates' ability to find employment. The study's findings allow for the drawing of a conclusion that might further the general discussion of IT.

A vital theme links with a thorough analysis of information technology. The results showed that IT can be applied in a commercial setting and encompasses all forms of communication. All electronic data can be exchanged and stored by computers, nevertheless, because of their incredible capacity. Furthermore, it has been demonstrated that information technology is highly innovative, has a very bright future, and greatly simplifies life for all people. Thus, it is imperative that all

Fig. 6. Perceptions of information technology's impact on the economic growth.

Kurdistan Regional Government (KRG) university graduates use this chance to benefit from contemporary IT. As mentioned by Arai (2023), information technology is going to be around for a long time, and one of the most fascinating things about it is how fast it is developing. With the advent of the internet and smartphones, technology has completely changed the way we work and live, and it does not seem to be stopping anytime soon(Arai, 2023).

Another theme relates to information technology's influence on the growth of job creation. The results demonstrated that having a thorough understanding of IT and its components academically helps people locate appropriate occupations. More importantly, however, having knowledge of and credentials in information technology will be very beneficial to people looking to work and make a livelihood. As a result, it's critical that graduates of KRG universities pursue further education after receiving their degrees, particularly by enrolling in computer-related vocational courses and programs that will increase their knowledge and experience. By putting this technique into practice, they will have the opportunity to get employment quickly and efficiently. As explained through (Zhao et al., 2022), a promising new sector that has the potential to revolutionize the world economy is IT. New avenues for economic expansion in underdeveloped nations are being made possible by digital technologies; in addition, the development of the internet is a keystone and catalyst for increased global connectivity and economic revolution. The best example is Africa, which has been able to create the most jobs for university graduates through information technology while increasing annual GDP growth by about 1.5 percentage points and reducing the number of poverty heads by 0.7 percentage points per year(Wankuru, 2019). Additionally, (Georgescu Irina, 2021) suggest that the majority of the highly skilled labor force could potentially be employed in the IT sector. Therefore, it is clear that the country needs to invest in education and/or skill development if it hopes to attract young people to the lucrative job growth in the IT sector.

The last significant theme relates to how information technology has an impact on the economy's growth. The findings demonstrated that IT has actively acted to advance the nation's economic sector. Information technology is not only helping to increase employment among young people but is also becoming a valuable investment for economic expansion. Therefore, it will undoubtedly have a very significant effect on the economic growth of the Iraqi KR when university graduates are able to obtain employment through IT. In order to revitalize its economy and pay attention to information technology, the KRG has to have a well-organized program.

Concerning (Faucheux et al., 2010), technology is viewed as a growth driver for the economy. It has an impact on it in two ways: directly through the industries that produce ICT and directly through the industries that are users. All economic sectors now employ ICT to the point where, in developed economies, indirect productivity gains associated with digitization and how it is used are frequently considered as the key vector of growth. Additionally, an examination of the work by Oliner and Sickel (2000, 2002) for the United States and Cette, Mairesse, and Kocoglu (2002, 2004) for France reveals that the IT sector's contribution, which was 0.3% per year in the United States until the middle of 1995, has lately increased to 0.8%. In the same time frame, it increased in France from 0% to 6% annually. Consequently, these findings lead to a very important conclusion about the IT-using industries throughout economic. IT use appears to be more significant in the United States. The above authors concur that IT grew in the late 1990s and subsequently enhanced the contribution of economic growth (Oliner, 2000, Cette et al., 2002).

In conclusion, this study reveals that graduates of universities in the Iraqi Kurdistan Region can find job opportunities through information technology. However, IT has further enhanced economic growth, especially in the Iraqi Kurdistan Region, which is able to take advantage of modern information technology components. This study is limited to the participation of university graduates from the Iraqi Kurdistan Region.